# Deep-learning-based on-chip rapid spectral imaging with high spatial resolution

Jiawei Yang,[1,2] Kaiyu Cui,[1,2,*] Yidong Huang,[1,2,3] Wei Zhang,[1,2,3] Xue Feng,[1,2] and Fang Liu,[1,2]

[1]*Department of Electronic Engineering, Tsinghua University, Beijing 100084, China*

[2]*Beijing National Research Center for Information Science and Technology (BNRist), Tsinghua University, Beijing 100084, China*

[3]*Bejing Academy of Quantum Information Science, Beijing 100084, China*

*Corresponding author: kaiyucui@tsinghua.edu.cn*

## ABSTRACT

Spectral imaging extends the concept of traditional color cameras to capture images across multiple spectral channels and has broad application prospects. Conventional spectral cameras based on scanning methods suffer from low acquisition speed and large volume. On-chip computational spectral imaging based on metasurface filters provides a promising scheme for portable applications, but endures long computation time for point-by-point iterative spectral reconstruction and mosaic effect in the reconstructed spectral images. In this study, we demonstrated on-chip rapid spectral imaging eliminating the mosaic effect in the spectral image by deep-learning-based spectral data cube reconstruction. We experimentally achieved four orders of magnitude speed improvement than iterative spectral reconstruction and high fidelity of spectral reconstruction over 99% for a standard color board. In particular, we demonstrated video-rate spectral imaging for moving objects and outdoor driving scenes with good performance for recognizing metamerism, where the concolorous sky and white cars can be distinguished via their spectra, showing great potential for autonomous driving and other practical applications in the field of intelligent perception.

**KEYWORDS:** spectral imaging, deep learning, metasurface

## 1. INTRODUCTION

Spectral imaging technology aims to capture spectral information for each two-dimensional spatial point to form a spectral data cube. It has been applied in a broad range of fields such as remote sensing[1,2], precision agriculture[3], medical diagnostics[4,5], food inspection[6], environmental monitoring[7], art conservation[8,9], and astronomy[10]. Traditional spectral imagers rely on either spatial scanning, as with whiskbroom scanning[11] and pushbroom scanning[12], or spectral scanning, as with filter wheels[13] and tunable filters[14,15]. However, scanning methods suffer from low acquisition speed, thus not applicable for dynamic recording of moving targets. To overcome this limitation, snapshot spectral imaging methods[16] are explored. Early snapshot techniques, such as integral field spectrometry[17-19], multispectral beam splitting[20], and image-replicating imaging spectrometer[21], still depend on light splitting and have bulky optical systems. With the development of compressive sensing (CS)[22,23], the computational snapshot spectral imaging technique[24] has attracted growing research interest, which can be categorized into three groups: coded aperture methods, speckle-based methods, and spectral filter array methods. For coded aperture methods, the classical system is coded aperture snapshot spectral imager (CASSI)[25-28], which uses fixed masks and dispersive elements to implement band-wise modulation. CASSI is able to capture and reconstruct hyperspectral images rapidly with deep-learning techniques. However, the complicated optical components lead to large system volume, which cannot meet the growing demand for portable applications. Speckle-based methods[29,30] utilize the wavelength dependence of speckle from a scattering media to achieve spectral imaging. Although the systems can be compact, their spectral resolution is limited by the speckle correlation through wavelengths. The spectral filter array methods can be viewed as an extension of Bayer filters, which use a super-pixel containing a group of spectral filters for spectral recovery. Even though this class of methods has a compact device size and high spectral accuracy, there is the mosaic effect in the reconstructed spectral images, where the recovered spectra for the edge points are inaccurate. Recently, our group demonstrated a snapshot spectral imaging chip based on metasurface-based spectral filter arrays with ultra-high center-wavelength accuracy of 0.04 nm and spectral resolution of 0.8 nm[31]. Furthermore, the spectral resolution is improved to 0.5 nm using metasurfaces with freeform shaped meta-atoms in our latest work[32]. However, apart from the aforementioned mosaic effect, long computation time for the data cube reconstruction is still required due to the classic iterative CS algorithm used here, which limits its application in the mobile systems with speed requirements, such as pilotless automobile.

In this work, we demonstrate on-chip mosaic-free rapid spectral imaging by employing advanced deep-learning-empowered algorithms developed for CASSI to the metasurface spectral imager reported in our previous work[32]. The metasurface spectral imager produces different amplitude modulation patterns for different spectral bands, which plays the role of a fixed mask plus a disperser in CASSI. Specifically, the spectral imager is designed by integrating a metasurface layer composed of 360×440 metasurface units with freeform shaped meta-atoms, onto a CMOS image sensor (CIS). There are totally 400 kinds of metasurface units, each of which has a distinctive spectral response function. As a proof of principle, we select 256×256 metasurface units and 26 wavelength bands from 450 to 700 nm with a step of 10 nm. A deep unfolding network based on the alternating direction method of multipliers (ADMM) algorithm, which is called ADMM-net[33], is adopted here for the fast reconstruction of spectral images. Here, the network is trained on a synthetic dataset containing 750,000 spectral data cubes with a size of 256×256×26 generated from the CAVE[34] and KAIST[35] datasets. Besides, we impose additive white Gaussian noise on the measurements to mimic real-test cases. On-chip rapid spectral imaging eliminating the mosaic effect is realized by applying the ADMM-net to reconstruct the spectral data cube directly. Compared with the point-by-point iterative spectral reconstruction, the ADMM-net achieves four orders of magnitude speed improvement, which enables a spectral data cube reconstruction rate of 55 per second, and the average spectral reconstruction fidelity exceeds 99% for a 24-patch Macbeth color checker. In practice, we demonstrate video-rate spectral image reconstruction for moving objects and outdoor driving scenes. It is found that the concolorous sky and moving white cars can be effectively distinguished by their spectra, while the existing driverless vehicles can easily mistake a white truck for the sky and cause a crash. Our approach can solve the huge safety problem caused by the defect in metamerism recognition for not only autonomous driving[36] but also other fields of intelligent perception, and shows great potential for various applications.

## 2. HARDWARE STRUCTURE

The schematic of the metasurface-based spectral imager is shown in Fig. 1a, as reported in Ref. [32]. A silicon metasurface layer containing 360×440 metasurface units is integrated on the image sensor with a microlens layer. Each metasurface unit is a periodic array with freeform shaped meta-atoms. There are 400 kinds of metasurface units with different periods and shapes, thus having distinctive transmission spectra. Therefore, the spectrum of incident light at the $(i,j)$ point can be reconstructed from the $N$ detected signals at surrounding points by solving such a system of linear equations:

$$\begin{bmatrix} y^{[i-n,j-n]} \\ \vdots \\ y^{[i+n,j+n]} \end{bmatrix} = \begin{bmatrix} M_1^{[i-n,j-n]} & M_2^{[i-n,j-n]} & \cdots & M_{N_\lambda}^{[i-n,j-n]} \\ \vdots & \vdots & \ddots & \vdots \\ M_1^{[i+n,j+n]} & M_2^{[i+n,j+n]} & \cdots & M_{N_\lambda}^{[i+n,j+n]} \end{bmatrix} \begin{bmatrix} x_1 \\ x_2 \\ \vdots \\ x_{N_\lambda} \end{bmatrix} \quad (1)$$

where $N = (2n+1)^2$ (typically, we set $n = 2$), $y^{[i,j]}$ denotes the measured signal at the $(i,j)$ point, $M_k^{[i,j]}$ represents the modulation intensity (i.e., transmittance of the metasurface unit) at the $(i,j)$ point for the $k$-th spectral channel, $x_k$ is the $k$-th element of the target spectrum vector at the $(i,j)$ point. In order to reconstruct the whole spectral data cube, we need to solve $N_x N_y$ groups of equations like Eq. (1), which is time-consuming. **For the iterative CS algorithm, we make an assumption that the spectra of incident light at the $N$ points around $(i,j)$ are the same in Eq. (1), which results in mosaic effect in the reconstructed spectral images.** In order to address the above problems, we propose to exploit the deep-learning algorithms to directly reconstruct the data cube inspired by CASSI[37] (see Supplement 1, S1 for details), as indicated in Fig. 1b.

The basic principle of CASSI is presented in Fig. 1c, where the spectral data cube is spatially coded by a fixed physical mask (coded aperture), and then spectrally sheared by a dispersive element, and finally measured by a detector. Therefore, for CASSI, a fixed mask and a disperser are used to achieve different masks at different spectral channels. In this work, the metasurface layer is used to achieve different masks at different wavelengths, and its mathematical model can be written as:

$$\begin{bmatrix} y^{[1,1]} \\ y^{[2,1]} \\ \vdots \\ y^{[N_x,N_y]} \end{bmatrix} = \begin{bmatrix} M_1^{[1,1]} & 0 & \cdots & 0 \\ 0 & M_1^{[2,1]} & \cdots & 0 \\ \vdots & \vdots & \ddots & \vdots \\ 0 & 0 & \cdots & M_1^{[N_x,N_y]} \end{bmatrix} \cdots \begin{bmatrix} M_{N_\lambda}^{[1,1]} & 0 & \cdots & 0 \\ 0 & M_{N_\lambda}^{[2,1]} & \cdots & 0 \\ \vdots & \vdots & \ddots & \vdots \\ 0 & 0 & \cdots & M_{N_\lambda}^{[N_x,N_y]} \end{bmatrix} \begin{bmatrix} x_1 \\ x_2 \\ \vdots \\ x_{N_\lambda} \end{bmatrix} \quad (2)$$

where $\boldsymbol{x}_k = \begin{bmatrix} x_k^{[1,1]} & x_k^{[2,1]} & \cdots & x_k^{[N_x,N_y]} \end{bmatrix}^T$ is the $k$-th frame of the spectral data cube. Taking the measurement noise into consideration, the Eq. (2) can be expressed in a vectorized formulation as follows:

$$\boldsymbol{y} = \boldsymbol{\Phi}\boldsymbol{x} + \boldsymbol{e} \quad (3)$$

where $\boldsymbol{y}$ is the compressed measurement, $\boldsymbol{x}$ is the target signal, $\boldsymbol{\Phi}$ is the sensing matrix and $\boldsymbol{e}$ is the measurement noise. As a proof of principle, we select 256×256 metasurface units for spectral imaging with 26 wavelengths from 450 nm to 700 nm ($N_x = N_y = 256, N_\lambda = 26$).

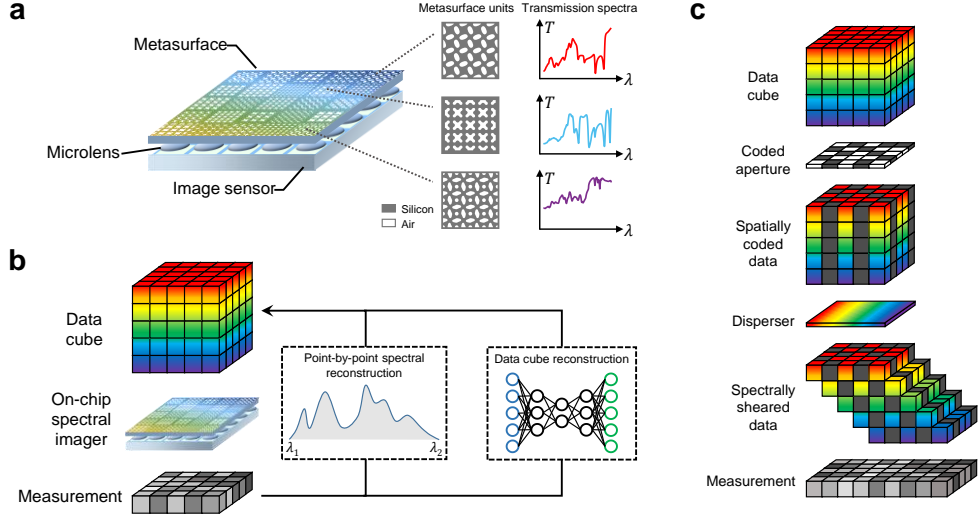

**Fig. 1** a Schematic representation of the metasurface-based spectral imager, which consists of a metasurface layer, a microlens layer and an image sensor layer. The silicon metasurface contains 360×440 metasurface units with freeform shaped meta-atoms. There are 400 types of metasurface units with distinctive transmission spectra. Typically, 5×5 metasurface units are combined to form a micro-spectrometer (spectral pixel). b Two methods of data cube reconstruction from the measurement of the spectral imager in a including the point-by-point spectral reconstruction using iterative optimization algorithms, and fast reconstruction of the whole data cube via deep learning algorithms. c Schematic diagram of coded aperture snapshot spectral imaging. The spectral data cube is first modulated by a fixed coded aperture (mask), then sheared by a disperser, and finally measured by a detector.

## 3. RECONSTRUCTION NETWORK

Here, we employ a deep unfolding network base on the ADMM algorithm dubbed ADMM-net for data cube reconstruction. The framework of ADMM-net with $K$ stages ($K$=12) is depicted in Fig. 2a. As in Ref. [33], let $v$ denote an estimate of the desired signal and by introducing two auxiliary variables $x, u$, the three steps for updating variables in each stage are listed as:

$$x^{(k)} = \left[\Phi^T \Phi + \gamma^{(k)} I\right]^{-1} \left[\Phi^T y + \left(v^{(k-1)} + u^{(k-1)}\right)\right] \quad (4)$$

$$v^{(k)} = \mathcal{D}_k\left(x^{(k)} - u^{(k-1)}\right) \quad (5)$$

$$u^{(k)} = u^{(k-1)} - \left(x^{(k)} - v^{(k)}\right) \quad (6)$$

Here, Eq. (4) represents a linear projection and can be solved in one shot[38]. Eq. (5) is a denoising process performed by the *CNN* as shown in Fig. 2a. We use a 15-layer *U*-net[39] as the denoiser, the architecture of which is described in Fig. 2b.

We construct a basic dataset containing 262 scenes with a size of 512×512×26 from the publicly available hyperspectral dataset CAVE and KAIST (see Supplement 1, S2 for details). The RGB images of selected 15 scenes in the basic dataset, which are converted from the spectral images via the International Commission on Illumination color-matching function[40], are shown in Fig. 2c. For model training, we randomly select 252 scenes from the basic dataset, and implement data augmentation to obtain 750,000 samples with a size of 256×256×26. The operations of data augmentation include random cropping, rotation, and multiplying with the spectra of LED light or sunlight sources (see Supplement 1, S3 for details). The remaining 10 scenes are downsampled to the size of 256×256×26 for testing. Besides, the loss function is the root mean square error between the ground truth and the output result of the network. The network is trained by the Adam optimizer on Pytorch using NVIDIA GeForce RTX 3080 GPUs. The total number of epochs is 300, and the batch size is set to 4 due to the limitation of GPU memory. The learning rate is set to 0.001 initially, and scales to 90% of the previous one every 20 epochs.

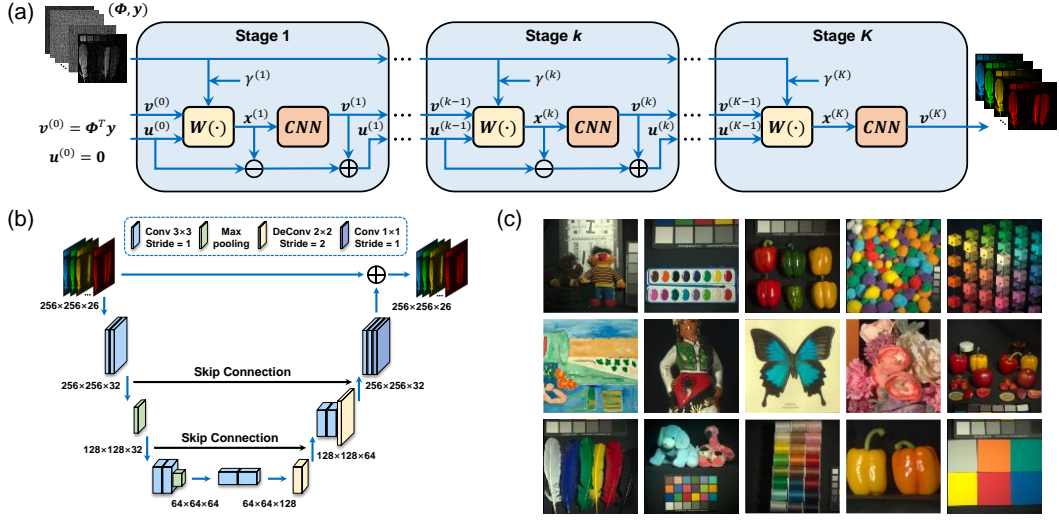

**Fig. 2** a Data flow graph of the ADMM-net with *K* stages, where each stage contains a linear projection $W(\cdot)$ denoting the computation in Eq. (4) and a *CNN* denoiser. b Architecture of the *U*-net used in a for denoising. c RGB images of 15 samples in the basic spectral dataset.

First, we conduct network training and test under noise-free and noisy conditions, respectively. For noisy conditions, we assume that each element of the measurement noise vector $e$ in Eq. (3) follows an independent zero-mean Gaussian distribution[38], that is, $e_i \sim \mathcal{N}(0, \sigma_n^2)$, where the standard deviation $\sigma_n$ is randomly chosen between 0 and 0.05 to increase the robustness to noise of different levels. The reconstruction results of two test scenes are given in Fig. 3. For the *crayon* scene without noise, it can be seen that the average fidelity of the recovered spectra for the selected three points exceeds 99.98%, as indicated in the legends of Fig. 3a. Under noisy conditions, we can see the ADMM-net can still recover the spectra reliably with an average fidelity of over 99.95%. Here the fidelity is defined as:

$$F(f_1, f_2) = \langle f_1, f_2 \rangle \quad (7)$$

where $f_1$, $f_2$ are the $l_2$-normalized ground truth and reconstructed result, respectively, and <> means the inner product. The four reconstructed exemplar frames of the *crayon* hyperspectral data are shown in Fig. 3b, which are highly consistent with the ground truth in both noise-free and noisy cases. For the *toy* scene without noise, the ADMM-net can also provide accurate reconstructed results as shown in Fig. 3c and Fig. 3d. In the noisy case, the spectral reconstruction quality is degraded slightly compared to the *crayon* scene since there are more fine spatial details in Fig. 3c, and the four frames in Fig. 3d are still recovered with high quality.

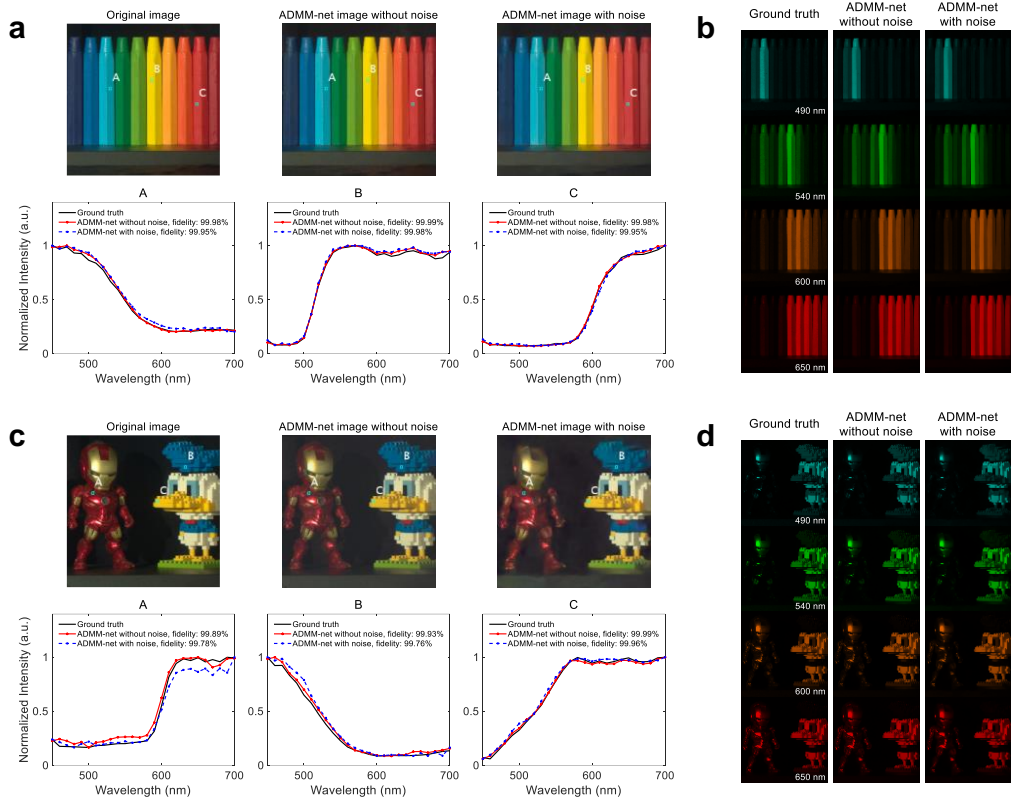

**Fig. 3** ac Reconstruction results of simulated *crayon* and *toy* hyperspectral data without and with noise, respectively. The original and reconstructed RGB images of the scene are shown on the top row with a size of 256×256 pixels. The spectra of three selected points are shown on the bottom row, where the fidelity of the reconstructed spectra is shown in the legends. bd Reconstructed frames of simulated *crayon* and *toy* hyperspectral data without and with noise, respectively. The noise level is randomly chosen from 0% to 5%.

## 4. RAPID SPECTRAL IMAGING

Then, we apply the ADMM-net trained with noise to reconstruct the real data from snapshot measurements captured by the spectral imager as shown in Fig. 4a. The metasurface layer with a size of 8 mm×6.4 mm is integrated on top of a CIS (Thorlabs, CS235MU). The spectral imager is assembled with a lens with a fixed focal length of 50 mm (Thorlabs, MVL50M23) for imaging. The recovered results for a standard 24-patch Macbeth color checker are displayed in Fig. 4b. To better show the reconstruction performance, we use a commercial spectral camera (Dualix Instruments, GaiaField Pro V10) based on line scanning to capture the spectral image as a reference. Moreover, we compare the results of ADMM-net with the results obtained by point-by-point spectral reconstruction using the iterative CS algorithm implemented by CVX[41], an open-source package for convex optimization. Here, we use 25 metasurface units for iterative spectral reconstruction by solving Eq. (1) with 601 wavelength channels (from 450 to 750 nm with a step of 0.5 nm), and then downsample the result to 26 channels (from 450 to 700 nm with a step of 10 nm). It can be clearly seen that ADMM-net outperforms the CVX in both spatial details and spectral accuracy. For the selected four points, the average fidelity of the reconstructed spectra using ADMM-net is 99.53% while the average fidelity for CVX is merely 97.32%. Additional reconstruction results for a Thorlabs box are provided in Supplement 1, S4.

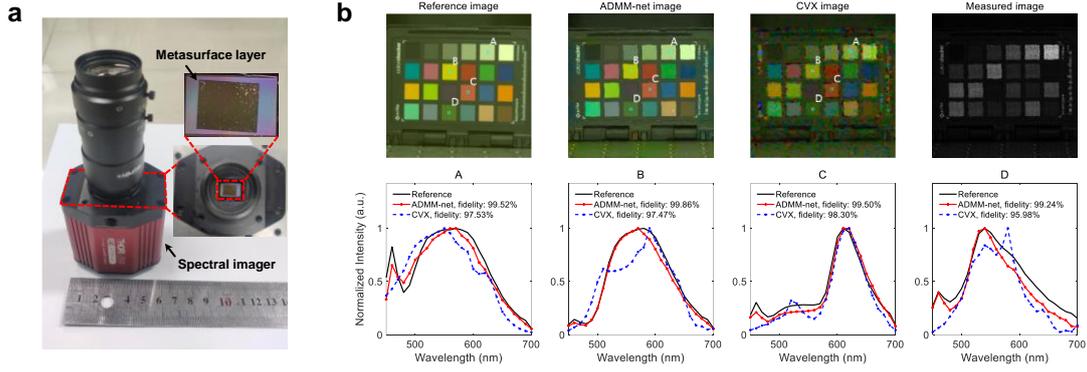

**Fig. 4** a The spectral imager with a metasurface layer (8×6.4 mm$^2$) integrated on a CMOS image sensor (Thorlabs, CS235MU) and a lens with fixed focal length (Thorlabs, MVL50M23) are used for spectral imaging. b Reconstruction results of real hyperspectral data of a 24-patch Macbeth color chart. The reference RGB image captured by a commercial spectral camera (Dualix Instruments, GaiaField Pro V10), the reconstructed RGB images using ADMM-net and CVX, and the snapshot measurement are shown on the top row from left to right with a size of 256×256 pixels. The spectra of four selected points are shown on the bottom row, where the fidelity of the reconstructed spectra is shown in the legends.

We compare the running time using different methods to obtain a spectral data cube, as reported in Table 1. For the commercial spectral camera based on line scanning, it spends about 1 minute to capture a data cube with a size of 256×256×26. For the computational methods, the ADMM-net only needs 1.72 seconds on CPU (Intel Xeon Gold 6226R) or 0.018 seconds on GPU (NVIDIA GeForce RTX 3080) to recover the spectral data cube of size 256×256×26, while the CVX takes 7767 or 4854 seconds for the spectral cube of size 256×256×601 or 256×256×26, respectively. In addition, we divide the running time by the number of wavelength channels as indicated in the last row of Table 1. It can be seen that the reconstruction speed of ADMM-net is over four orders of magnitude faster than CVX, which enables 256×256×26×55 4D data cube reconstruction per second for rapid spectral imaging in real applications.

**Table 1** Comparison of different methods

| Methods | Line scanning | ADMM-net | CVX | |
|---|---|---|---|---|
| Data cube size | 256×256×26 | 256×256×26 | 256×256×601 | 256×256×26 |
| Running time (s) | ~60 | 1.72 @CPU (Intel Xeon Gold 6226R) <br> 0.018 @GPU (NVIDIA GeForce RTX 3080) | 7767 @CPU | 4854 @CPU |
| Running time per channel (s) | ~2.31 | 0.066 @CPU <br> 0.00069 @GPU | 12.92 @CPU | 186.69 @CPU |

To show the ability of high-speed spectral image reconstruction using our approach, we conduct experiments of video spectral imaging. The first example is a moving Thorlabs box indoors with an LED light source. The reconstructed hyperspectral video contains 200 frames totally (19.84 seconds), from which 10 frames are extracted and shown in Fig. 5. From the results of recovered RGB images and spectral images at four wavelengths, we can see the spatial, spectral and motion details are reconstructed with high quality. The full video is provided in the Supplement 2.

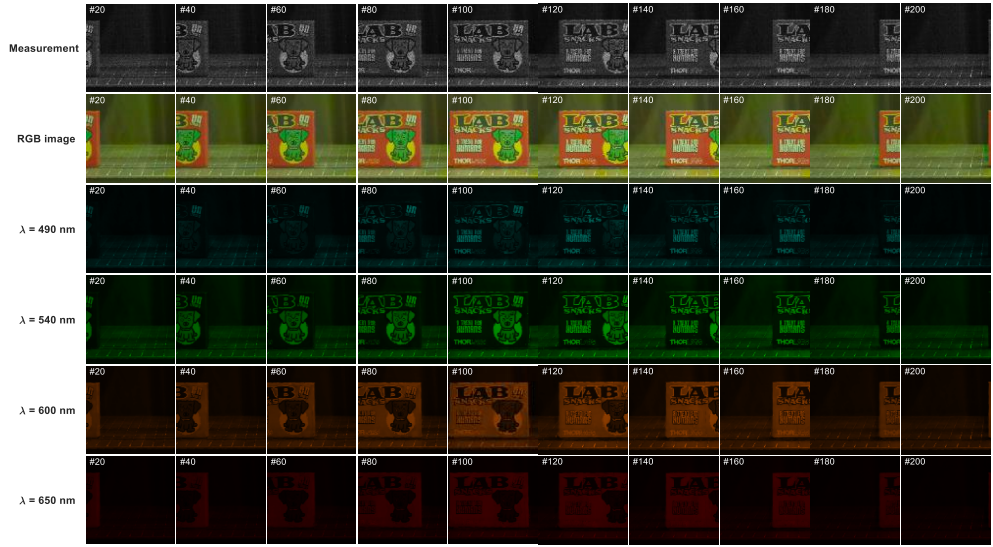

**Fig. 5** Reconstructed results of real hyperspectral data of a moving Thorlabs box. Snapshot measurements captured by the on-chip spectral imager are shown on the first row. The reconstructed RGB images are shown on the second row. The reconstructed spectral images at different wavelengths are shown on the third to sixth rows.

The second example is outdoor driving scene under the sunlight. The reconstructed video contains 300 frames (8.38 seconds) with a shorter exposure time compared with the first example. That is, our approach can support a refresh rate of about 2 data cubes per meter for autonomous vehicles running at 60km/h. Note that the refresh rate contains both the capture and reconstruction process while the aforementioned 55 spectral data cube per second only considers the reconstruction process. Figure 6 shows the recovered RGB images and spectral images at three wavelengths of the extracted 8 frames, as well as the recovered spectra for the selected three points. It can be seen that the driving cars with different colors can be reconstructed with fine spatial details, and the spectral characteristics are also recovered with high quality. In particular, from the spectra of points A and B in the frame 20 and 100, we can see that there are obvious differences between the spectra of the sky and white car. Hence, the sky and white cars can be distinguished using our approach, which is significant for autonomous driving with the defect in metamerism recognition as a huge security concern[36]. The full video of this example is provided in the Supplement 3.

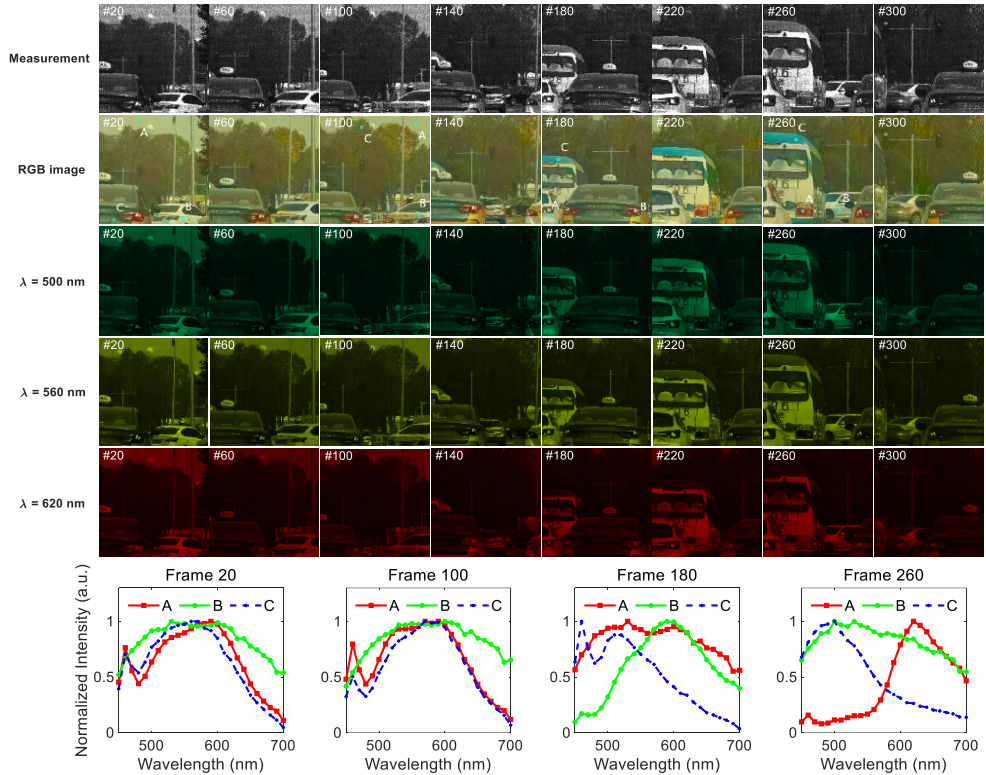

**Fig. 6** Reconstructed results of real hyperspectral data of an outdoor driving scene. Snapshot measurements captured by the on-chip spectral imager are shown on the first row. The reconstructed RGB images are shown on the second row. The reconstructed spectra of three selected points at different time frames are shown on the third row.

## 5. DISCUSSION

There are some aspects remained to be improved in this work. First, the wavelength sampling interval is set to 10 nm due to the limitation of GPU memory. Reconstruction of spectral images with more wavelength bands can be anticipated with improved computational power and GPU memory. Second, the capacity of the training dataset is still insufficient. Spectral datasets with large scale, accuracy and diversity are required to improve the network performance. Third, the structure of the ADMM-net can be further improved such as increasing the number of stages, using deeper and more advanced CNN denoisers. Besides, we only consider Gaussian white noise to simulate the real situations in the network training. A more complete and accurate noise model taking both Gaussian white noise and shot noise into account is required for better performance.

## 6. CONCLUSIONS

In summary, we proposed to employ a deep unfolding network called ADMM-net for fast spectral image reconstruction from the snapshot measurement of the metasurface-based spectral imager. Our approach shows excellent performance in both simulated and real data reconstruction. Notably, we achieved four orders of magnitude improvement in reconstruction speed with high spectral accuracy and without mosaic effect in real experiments, compared with conventional point-by-point iterative spectral reconstruction. Moreover, we demonstrate video-rate spectral imaging for moving objects and outdoor driving scenes with good performance of metamerism recognition, where the concolorous sky and white cars can be effectively distinguished according to their spectra. Our approach can provide real-time capture and reconstruction of hyperspectral images, paving the way for autonomous driving and other various real-time applications in the field of intelligent perception.

## METHODS

The metasurface-based spectral imager is fabricated on a Silicon-On-Insulator (SOI) wafer with a 220-nm silicon layer. First, the metasurface patterns are defined via electron beam lithography (EBL) using ZEP resist, and transferred onto the top silicon layer through reactive ion etching (RIE). Then, the silicon dioxide under the patterned area is removed through immersion in buffered hydrofluoric (HF) acid in a water bath, at 40 °C, for approximately 3 min. Finally, the suspended silicon metasurface layer is peeled off and transferred onto a CIS chip via a polydimethylsiloxane (PDMS) adhesion layer.

## REFERENCES

1. Shaw, G. A. et al. Spectral imaging for remote sensing. *Lincoln laboratory journal* **14**, 3-28(2003).
2. Williams, L. J. et al. Remote spectral detection of biodiversity effects on forest biomass. *Nat. Ecol. Evol.* **5**, 46-54(2021). https://doi.org/10.1038/s41559-020-01329-4
3. Lebourgeois, V. et al. Can commercial digital cameras be used as multispectral sensors? A crop monitoring test. *Sensors* **8**, 7300-7322(2008). https://doi.org/10.3390/s8117300
4. Lu, G., Fei, B. Medical hyperspectral imaging: a review. *J Biomed Opt* **19**, 010901(2014). https://doi.org/10.1117/1.JBO.19.1.010901
5. Yao, L. et al. Image enhancement based on in vivo hyperspectral gastroscopic images: a case study. *J Biomed Opt* **21**, 101412(2016). https://doi.org/10.1117/1.JBO.21.10.101412
6. Feng, Y. Z., Sun, D. W. Application of hyperspectral imaging in food safety inspection and control: a review. *Crit Rev Food Sci Nutr* **52**, 1039-1058(2012). https://doi.org/10.1080/10408398.2011.651542
7. Stuart, M. B., McGonigle, A. J. S., Willmott, J. R. Hyperspectral imaging in environmental monitoring: a review of recent developments and technological advances in compact field deployable systems. *Sensors* **19**, 3071(2019). https://doi.org/10.3390/s19143071
8. Liang, H. Advances in multispectral and hyperspectral imaging for archaeology and art conservation. *Appl. Phys. A* **106**, 309-323(2012). https://doi.org/10.1007/s00339-011-6689-1
9. Gabrieli, F. et al. Near-UV to mid-IR reflectance imaging spectroscopy of paintings on the macroscale. *Sci. Adv.* **5**, eaaw7794(2019). https://doi.org/10.1126/sciadv.aaw7794
10. Bahauddin, S. M., Bradshaw, S. J., Winebarger, A. R. The origin of reconnection-mediated transient brightenings in the solar transition region. *Nat. Astron.* **5**, 237-245(2021). https://doi.org/10.1038/s41550-020-01263-2
11. Green, R. O. et al. Imaging spectroscopy and the airborne visible/infrared imaging spectrometer (AVIRIS). *Remote Sens Environ* **65**, 227-248(1998). https://doi.org/10.1016/S0034-4257(98)00064-9
12. Mouroulis, P., Green, R. O., Chrien, T. G. Design of pushbroom imaging spectrometers for optimum recovery of spectroscopic and spatial information. *Appl. Opt.* **39**, 2210-2220(2000). https://doi.org/10.1364/AO.39.002210
13. Zhang, C. et al. A novel 3D multispectral vision system based on filter wheel cameras. In *Proceedings of the 2016 IEEE International Conference on Imaging Systems and Techniques (IST)* (pp. 267-272)(2016). https://doi.org/10.1109/IST.2016.7738235
14. Gat, N. Imaging spectroscopy using tunable filters: a review. *Wavelet Applications VII* **4056**, 50-64(2000). https://doi.org/10.1117/12.381686
15. Antila, J. et al. Spectral imaging device based on a tuneable MEMS Fabry-Perot interferometer. In *Conference on Next-Generation Spectroscopic Technologies V* **8374**, 83740F(2012). https://doi.org/10.1117/12.919271

16. Hagen, N. A., Kudenov, M. W. Review of snapshot spectral imaging technologies. *Opt. Eng.* **52**, 090901(2013).

17. Bowen, I. S. The Image-Slicer a Device for Reducing Loss of Light at Slit of Stellar Spectrograph. *The Astrophysical Journal* **88**, 113(1938). https://doi.org/10.1086/143964

18. Gat, N. et al. Development of four-dimensional imaging spectrometers (4D-IS). In *Conference on Imaging Spectrometry XI* **6302**, 63020M(2006). https://doi.org/10.1117/12.678082

19. Bacon, R. et al. The integral field spectrograph TIGER. *Very Large Telescopes and their Instrumentation*, Vol. 2 **30**, 1185(1988).

20. Stoffels, J., Bluekens, A. A. J., Jacobus, M. P. P. Color splitting prism assembly. *U.S. Patent* 4,084,180(1978).

21. Harvey, A. R., Fletcher-Holmes, D. W. High-throughput snapshot spectral imaging in two dimensions. In *Conference on Spectral Imaging - Instrumentation, Applications and Analysis II* **4959**, 46-54(2003). https://doi.org/10.1117/12.485557

22. Donoho, D. L. Compressed sensing. *IEEE Trans. Inf. Theory* **52**, 1289-1306(2006). https://doi.org/10.1109/TIT.2006.871582

23. Candès, E. J., Romberg, J., Tao, T. Robust uncertainty principles: Exact signal reconstruction from highly incomplete frequency information. *IEEE Trans. Inf. Theory* **52**, 489-509(2006). https://doi.org/10.1109/TIT.2005.862083

24. Huang, L. et al. Spectral imaging with deep learning. *Light Sci. Appl.* **11**, 1-19(2022). https://doi.org/10.1038/s41377-022-00743-6

25. Gehm, M. E. et al. Single-shot compressive spectral imaging with a dual-disperser architecture. *Opt. Express* **15**, 14013-14027(2007). https://doi.org/10.1364/OE.15.014013

26. Wagadarikar, A. et al. Single disperser design for coded aperture snapshot spectral imaging. *Appl. Opt.* **47**, B44-B51(2008). https://doi.org/10.1364/AO.47.000B44

27. Correa, C. V., Arguello, H., Arce, G. R. Snapshot colored compressive spectral imager. *JOSA A* **32**, 1754-1763(2015). https://doi.org/10.1364/JOSAA.32.001754

28. Lin, X. et al. Spatial-spectral encoded compressive hyperspectral imaging. *ACM Trans. Graph.* **33**, 1-11(2014). https://doi.org/10.1145/2661229.2661262

29. Sahoo, S. K., Tang, D., Dang, C. Single-shot multispectral imaging with a monochromatic camera. *Optica* **4**, 1209-1213(2017). https://doi.org/10.1364/OPTICA.4.001209

30. French, R., Gigan, S., Muskens, O. L. Speckle-based hyperspectral imaging combining multiple scattering and compressive sensing in nanowire mats. *Opt. Lett.* **42**, 1820-1823(2017). https://doi.org/10.5258/SOTON/D0006

31. Xiong, J. et al. Dynamic brain spectrum acquired by a real-time ultraspectral imaging chip with reconfigurable metasurfaces. *Optica* **9**, 461-468(2022). https://doi.org/10.1364/OPTICA.440013

32. Yang, J. et al. Ultraspectral Imaging Based on Metasurfaces with Freeform Shaped Meta-Atoms. *Laser Photonics Rev.* 2100663(2022). https://doi.org/10.1002/lpor.202100663

33. Meng, Z., Jalali, S., Yuan, X. Gap-net for snapshot compressive imaging. arXiv preprint arXiv:2012.08364(2020).

34. Yasuma, F. et al. Generalized assorted pixel camera: postcapture control of resolution, dynamic range, and spectrum. *IEEE Trans Image Process* **19**, 2241-2253(2010). https://doi.org/10.1109/TIP.2010.2046811

35. Choi, I. et al. High-quality hyperspectral reconstruction using a spectral prior. *ACM Trans. Graph.* **36**, 218(2017). https://doi.org/10.1145/3130800.3130810

36. Liang, J. et al. Material based salient object detection from hyperspectral images. *Pattern Recognit* **76**, 476-490(2018). https://doi.org/10.1016/j.patcog.2017.11.024

37. Yuan, X., Brady, D. J., Katsaggelos, A. K. Snapshot compressive imaging: Theory, algorithms, and applications. *IEEE Signal Process Mag* **38**, 65-88(2021). https://doi.org/10.1109/MSP.2020.3023869

38. Liu, Y. et al. Rank minimization for snapshot compressive imaging. IEEE Trans. *Pattern Anal. Mach. Intell.* **41**, 2990-3006(2018). https://doi.org/10.1109/TPAMI.2018.2873587

39. Ronneberger, O., Fischer, P., Brox, T. U-net: Convolutional networks for biomedical image segmentation. In *the 18th International Conference on Medical Image Computing and Computer-Assisted Intervention (MICCAI)* (pp. 234-241)(2015). https://doi.org/10.1007/978-3-319-24574-4_28
40. Smith, T., Guild, J. The CIE colorimetric standards and their use. *Trans. Opt. Soc.* **33**, 73(1931). https://doi.org/10.1088/1475-4878/33/3/301
41. M. Grant and S. Boyd, CVX: Matlab software for disciplined convex programming, version 2.0 beta. http://cvxr.com/cvx, (2013).

## MISCELLANEA

**Availability of data and materials**

The data and materials that support the findings of this study and custom codes are available from the corresponding author upon reasonable request.

**Competing interests**

The authors declare no conflicts of interest.

**Funding**

The National Key Research and Development Program of China (2018YFB2200402); Beijing Municipal Science and Technology Commission (Z201100004020010); Beijing Frontier Science Center for Quantum Information; and Beijing Academy of Quantum Information Sciences.

**Authors' contributions**

J.Y. conceived the study and wrote the manuscript. K.C. and Y.H. supervised the project, provided much support on experimentation, and reviewed the manuscript with contributions from all other co-authors. All authors read and approved the manuscript.

**Acknowledgments**

The authors would like to thank Beijing Seetrum Technology Co., Ltd. for their valuable discussions, and Tianjin H-Chip Technology Group Corporation, Innovation Center of Advanced Optoelectronic Chip and Institute for Electronics and Information Technology in Tianjin, Tsinghua University for their fabrication support with SEM and ICP etching.